\documentclass[12pt]{article}
\usepackage{epsfig}

\def\beq{\begin{equation}}
\def\eeq#1{\label{#1}\end{equation}}
\def\eeqn{\end{equation}}

\def\beqa{\begin{eqnarray}}
\def\eeqa#1{\label{#1}\end{eqnarray}}
\def\eeqan{\end{eqnarray}}

\let\bar=\overbar

\def\Dslash{\not{\hbox{\kern-4pt $D$}}}
\def\dslash{\not{\hbox{\kern-2pt $\del$}}}

\def\msb{{\bar{\ssstyle M \kern -1pt S}}}

\def\Title#1{\begin{center} {\Large {\bf #1} } \end{center}}

\begin{document}

\Title{Physics of Gamma-Ray Bursts: Turbulence, Energy Transfer and
Reconnection}

\bigskip\bigskip


\begin{raggedright}

{\it A. Lazarian (UW-Madison)\index{Lazarian, A.}
V. Petrosian (Stanford), H. Yan (UW-Madison) {\rm \&} J. Cho (UW-Madison) }
\bigskip\bigskip
\end{raggedright}

\section{Introduction}

Understanding of the nature of gamma-ray bursts (GRBs) is one of the
challenging problem facing the astrophysics community. The field of
gamma ray bursts is a rapidly developing one and we expect that new 
missions, like SWIFT and GLAST will bring the field
to a new quantitative level.

As a result of the intensive observational and theoretical efforts a basic
picture of the gamma-ray bursts  has emerged. The long standing question about
the distances of these sources has been resolved and
their cosmological origin is by now well established.
As has been known for some times now, this combined with their short time
scales, point to the relativistic fireball (or jet) as the most likely model
(see e.g.  \cite{MRee1,MRee2}).  Such fireballs could arise from
rapid and episodic accretion onto a stellar size black hole produced by
either a
hypernova-collapsar (\cite{Pacz, Woosley96}) or mergers of compact
objects.  The investigations of the spectra of both the prompt gamma-rays and
the afterglows seem to point to the synchrotron process as the most likely
source of the radiation (\cite{Tava,Katz,Sari}) but other models involving 
Compton scattering have also been proposed.
These models require presence of  relativistic electrons and either
strong magnetic fields or a source of soft photons.  The particle acceleration
takes place either in internal shocks arising from the episodic nature of the
accretion, for the gamma-ray emission, or in an external shock arising from
interaction of the fireball with the surrounding medium,
for the afterglows (\cite{Sari,Chiang}).  Some of these
aspects of GRBs have received some scrutiny but most of
them are still at the stage of a back-of-an-envelop calculations.
The detail of the explosion, the formation of the fireball, its 
propagation, the generation of
the shocks, the source of the magnetic fields and soft photons, the particle
acceleration process, and the details of the radiation process are all still
outstanding questions.


The goal of this review is to attract the attention of the community to
a number of physical processes that seem to be very relevant for these sources.
The radiation process, the one most directly related to observations, 
has received a lot of attention.  The next step toward the building of a credible
model is the determination of the mechanism for the acceleration of nonthermal
relativistic particles.  Most current
studies of GRBs do not deal with the mechanism of the acceleration; the
common practice is to assume presence of an isotropic distribution of electrons
with a either a simple power law spectrum, 
$f(\gamma) \propto \gamma^{-\delta}$,
with the Lorentz factors in the range $\gamma_{\rm min}<\gamma<\gamma_{\rm
max}$, a broken power law, or one with an exponential cutoff.  More
importantly, the crucial question of how the energy of the fireball,
which is carried mainly by protons,
is transferred to the radiating electrons has not received adequate
attention. In this review we shall discuss the possibility that {\it
stochastic acceleration  by plasma turbulence} is the
agent for this energy transfer and for the acceleration of the electrons.
In the next section (\S 2) we shall present a simple model of the
gamma ray burst and in \S 3 we discuss the stochastic
mechanism for particle acceleration.
Recent {\it advances in understanding of the nature of turbulence}
provide solid foundations for addressing those questions. These results
are discussed in \S 4 and \S 5.
Finally, in view of the 
emerging interest in the role 
{\it magnetic reconnection}in GRBs,\footnote{This
surge of interest was quite clear during the conference.} we shall briefly
discuss how turbulence may make reconnection both fast and bursty (\S 6).
The mechanism of stochastic reconnection 
that we discuss releases most of its energy in the form of
turbulent motions. Therefore our scenario of stochastic acceleration
of electrons
should be relevant to both shock-induced fireball and to reconnection
mechanism. The summary is provided in \S 7.

\section{A Possible Scenario}


The large distances and rapid variability of GRBs require release of a large
amount of energy ${\cal E} > 10^{53}(\Omega_b/4\pi)$ ergs, where
$(\Omega_b/4\pi)$ is the beaming factor, in a small region and within a short
time.  This in turn indicates presence of a high energy density and a
relativistic outflow with a bulk Lorentz factor $\Gamma \sim 10^3$ involving a
baryon mass of $M={\cal E}/\Gamma c^2$.  The light curves of
GRBs show a varied structure sometimes
involving many pulses, indicating the
episodic nature of the process.  After the initial fireballs associated with
these episodes cool and become transparent, the result is a series of
relatively cold ``shells" of masses $M_i$ moving with
 Lorentz factors $\Gamma_i$.  It
is believed that the interaction of these shells with each other (faster ones
running into slower ones) and with the surrounding medium (interstellar or
preburst stellar wind) converts a large fraction of the kinetic energy 
$(\Gamma - 1)Mc^2$ into prompt gamma-rays and afterglow photons.
These interactions will
clearly produce shocks.  The shocks produced by shell-shell interactions are
referred to as internal and those produced by shell-medium interactions are
referred to as external shocks.  It is generally believed that the burst proper
is produced by the internal shocks and that the afterglow is produced by an
external shock (\cite{Rami}), although a model of gamma-ray
production via external shocks also exists (\cite{Dermer99}).

We could
envision the following model for production of the gamma-rays and afterglows.
The interaction of two shells with Lorentz factors
$\Gamma_1$ and $\Gamma_2$ (or
a shell and the external medium; $\Gamma_2=1$) gives rise to a
relativistic shock front.  (For a description of structure of
relativistic shocks see \cite{BM76}). In the frame of the shock the
particles from the slow shell (or the external medium) will enter the shocked
region, essentially as a monoenergetic beam, with a relative Lorentz factor
$\Gamma_{\rm rel}\simeq
(\Gamma_1/\Gamma_2+\Gamma_2/\Gamma_1)/2$.  This is subject to
the well known two stream instability and can give rise to plasma turbulence
which dissipates the bulk of the kinetic energy of the system.  Most of the
energy is transferred to MHD turbulent motions (Alfv\'en waves or fast modes) 
by the instability induced by the proton beam (\cite{Pohl}).  Fast modes and Alfv\'en 
waves in magnetically dominated plasma
form distinct weakly coupled cascades which transfer energy
to small scales and possibly  to whistler waves (see \cite{GS95,CL02b}).  
Transit time damping  and direct scattering by the whistler waves
can accelerate the electrons.  In this scenario the electrons behind the 
shock are continually being accelerated as they also radiate, or lose 
energy by other ways, until the shells are completely merged
(or stopped by the external medium) and the shocks are dissipated.

\section{Acceleration Mechanism}

The three acceleration processes most usually proposed in astrophysical
sources are acceleration by static electric fields parallel to the magnetic
field, first order Fermi acceleration by shocks, and stochastic or second 
order Fermi acceleration by turbulence (see reviews 
\cite{EH91,Blandford,Jones}).  Large scale {\bf Electric Fields} needed to 
accelerate
charged particles to significant energies can be maintained only if the
electrical resistivity is anomalously large (\cite{Tsu}) or if the plasma
density $n$ is high and the Dreicer field is large (\cite{Holman}).
This does not seem to be a natural mechanism for GRBs.

{\bf Shocks} are the most commonly considered mechanism of acceleration because
they can quickly accelerate particles to very high energies.  In a
nonrelativistic shock, acceleration to high energies takes place by repeated
passages of the particles across the shock.  This requires the existence of
some scattering agents.  The most likely agent for scattering charged 
particles is plasma turbulence.  Presence of turbulence in the downstream 
region is natural but in the upstream region is problematic;
self generation of turbulence by the accelerated particles is often assumed.  
In any case, the rate of energy gain is governed by this scattering
rate which (for charged particles tied to magnetic fields) is proportional to
the pitch angle, diffusion coefficient $D_{\mu\mu}$; $\mu$ is the cosine of 
the pitch angle.  For a relativistic shock of Lorentz factor 
$\Gamma\gg 1$ most of the energy gain(equal to
$\Gamma mc^2$) occurs at the first passage.  Subsequent crossings, if any,
increase this energy by a factor of at most a few.  It has been shown (e.g.
 \cite{Kirk}) that this results in a power law spectrum
with a well defined index, $\delta = -2.23$, which would give a synchrotron
photon index of -1.62 (or -2.12 for {\it cooling} spectra).  Early
observations of afterglows seem to suggest such values of the index.  However,
this is not the case for all afterglows and it certainly is not true for the
prompt gamma-rays,
where the spectra involve at least two indexes both of which show a wide dispersion extending from $>0$ to $< -3$
(see \cite{LP00}).  
In addition, shock acceleration cannot be the sole mechanism for
the GRBs for the following reason.

The turbulence needed for the scattering can also accelerate particles {\bf
stochastically} at a rate $D_{pp}/p^2$, where $D_{pp}$ is the momentum 
diffusion
coefficient.  In most studies of astrophysical sources, and at high energies,
the pitch angle scattering rate is higher than the momentum diffusion rate
($D_{\mu\mu} \gg D_{pp}/p^2$) so that shock acceleration is more efficient than
direct acceleration by turbulence.  However, {\bf stochastic acceleration} 
becomes more efficient when the above inequality is reversed, $D_{pp}/p^2
\gg D_{\mu\mu}$ (\cite{ParkP}), which occurs at low energies, but more 
importantly for our purpose here, is true for high $B$ and low $n$ plasmas, 
where the formally defined Alfv\'en velocity (in units of c) $\beta_A=B/(4\pi
nm_pc^2)^{1/2}$ exceeds unity.  In this case the electric field fluctuations
$\delta E \sim \beta_A \delta B$ are larger than the magnetic field
fluctuations $\delta B$, which means a faster acceleration 
($\propto \delta E^2$) than scattering rate ($\propto \delta B^2$).  
This trend can be seen in results published in \cite{HPetro, PryP}.
Thus, under these conditions, once the particle
crosses the shock front into the turbulent region behind the shock, it will
undergo stochastic acceleration
much faster than it can be turned around to cross the shock
again.

These conditions are what is present in GRBs.  The observed
values of photons energies require magnetic fields 
$B \geq 10^4$G.  Various arguments (see review \cite{Piran}) suggest that 
$n\leq 10^8$ cm$^{-3}$ and therefore $\beta_A\gg 1$.

\section{Magnetic Turbulence and Particle Acceleration}

The rate of stochastic acceleration of electrons depends on the properties of
magnetic turbulence. Almost everyone would agree that incorporation of 
this aspect is necessary for a
realistic model, but the complexity of the problem has made many researchers
wary of MHD turbulence and has resulted in a tendency to avoid dealing with the
phenomenon.  However, recently a substantial progress has been achieved in the
field enabling us to deal with this problem adequately.  First of all,  
simple
scaling model of incompressible MHD turbulence developed by Goldreich \&
Shridhar (\cite{GS95}) has been successfully tested and extended 
recently (see review \cite{CLV02b}).  Moreover, important advances in
understanding of MHD turbulence in compressible media (\cite{LG02,CL, CL03}) 
make an adequate quantitative
investigation of turbulence possible for the first time.


Turbulence is ubiquitous in astrophysics.  All turbulent systems have one 
thing in common.  They have a large ``Reynolds number", which is the ratio 
of the viscous drag time on the largest scales $L^2/\nu$ to the eddy 
turnover time of a parcel of gas $L/V$; $Re\equiv LV/\nu$, where $L$ 
is the characteristic scale or driving scale of the system, $V$ is the 
velocity difference over this scale, and
$\nu$=viscosity),.  A similar parameter, the ``magnetic Reynolds number",
$Rm$$\equiv LV/\eta$ (with $\eta$ as the magnetic diffusion) is the ratio 
of the magnetic
field decay time $L^2/\eta$ to the eddy turnover time $L/V$.  The properties
of the flows on all scales depend on $Re$ and $Rm$.  Flows with $Re<100$ are
laminar; chaotic structures develop gradually as $Re$ increases.   For the
gamma-ray bursts both Reynolds numbers are $>>>1$ and fluid motions are
expected to be extremely chaotic.

{\it Hydrodynamic turbulence} of an incompressible fluid (or Kolmogorov
turbulence) is the simplest example of turbulence.  For instance, an 
obstacle of size $L$ in a flow
excites motions on scales of the order $L$.  The turbulent energy
injected at this scale cascades to progressively smaller and smaller scales
at the eddy turnover rate, with negligible energy losses along the cascade.
Ultimately, the energy reaches the molecular dissipation scale $l_d$, i.e.  
the scale, where the local $Re\sim 1$, and is dissipated there.  The scales 
between
$L$ and $l_d$ are called the {\it inertial range} which typically covers many
decades.  The motions over the inertial range are {\it self similar} and this
provides tremendous simplification for theoretical description.

{\it MHD turbulence} is more complex because the frozen-in magnetic
fields alter substantially the dynamics of fluid.  GRBs happen in 
magnetically dominated plasma with
$\beta_p=4\pi P_{gas}/B^2 < 10^{-4}$, where $P_{gas}$ is the gas pressure.
Turbulence for a low beta regime has been studied in \cite{CL} (see Fig.~1).

\begin{figure}
\begin{center}
\epsfig{file=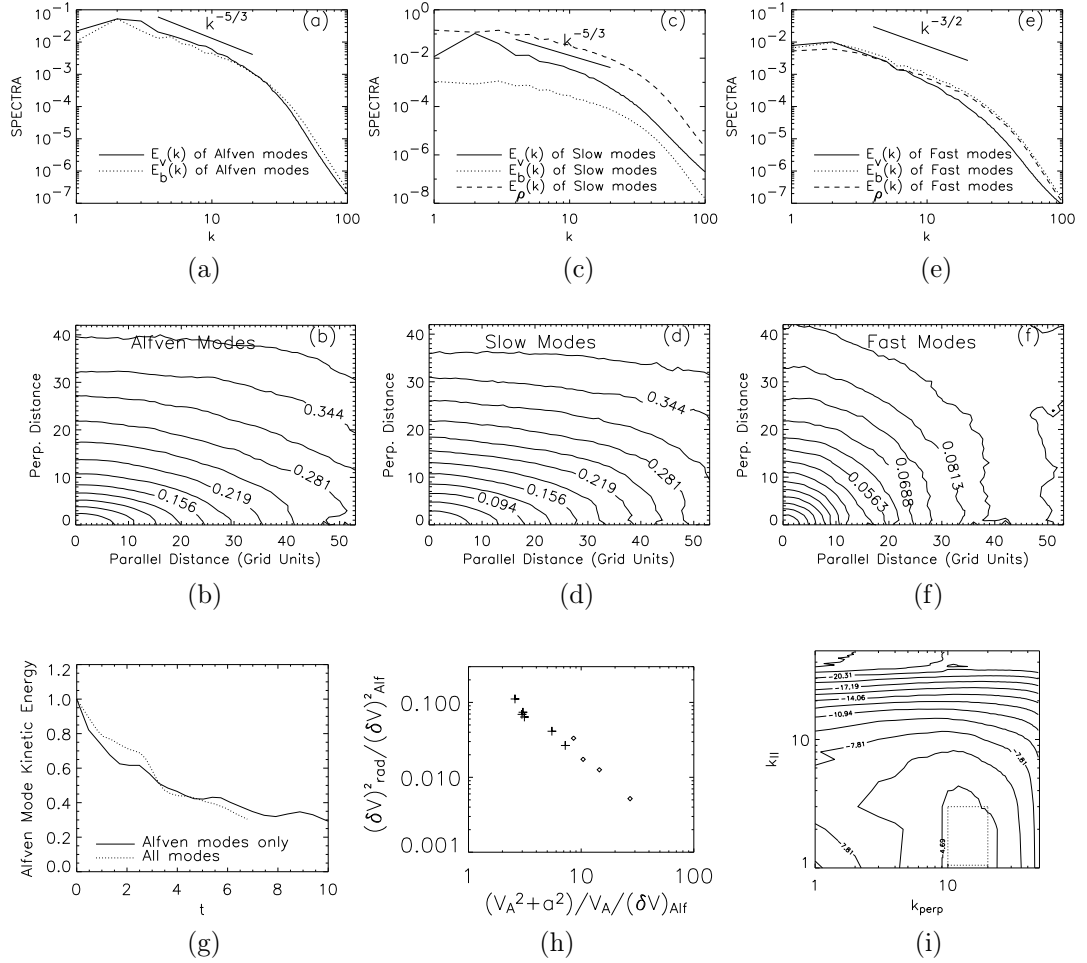,height=5 in} \hfil
\caption{\footnotesize
Statistics of MHD turbulence in low beta plasma 
(mostly from \cite{CL, CL03a}).
Results of driven compressible turbulence with Mach number (i.e
$V/V_s$, where $V_s$ is the sound velocity) equal to 2.2 and
the Alfv\'enic Mach number (i.e. $V/V_A$) 
equal to 0.7. The procedure of separation of various modes is explained
in \cite{CL}.  (a) The Alfv\'en modes show Kolmogorov-type scaling; (c)
Kolmogorov-type scaling is also true for slow modes; (e) Scaling of
fast modes is argued in \cite{CL} to follow the spectrum of accoustic
turbulence, i.e. $E(k)\sim k^{-3/2}$. While the differences in the 
slope of the fast, slow and Alfv\'en modes are not so different,
they exhibit very different anisotoropies. The isocontours of equal
correlation (measured with structure functions $\langle (v(\bf x_1)-
v(\bf x_2))^2\rangle$) are stretched along the direction of magnetic
field for slow and Alfv\'en modes (see (b) and (d)), but isotropic for 
fast modes (see (f)). By now similar results have been confirmed for
plasmas with different Mach numbers \cite{CL03} and ratios of gas to
magnetic pressure. 
         (g) Decay of Alfv\'en turbulence.
             The decay rate of Alfv\'en turbulence is not strongly affected by
             the presence of slow and fast modes.
             In the solid line, slow and fast modes are not present
             at the beginning of the simulation.
             In the datted line, we include slow and fast modes
             at the beginning. $\beta < 1$. 
           {}From Cho \& Lazarian (\cite{CL03}).
         (h) The ratio of $(\delta V)_f^2$ to
          $(\delta V)_A^2$.
          Initially, only Alfv\'en modes are present.
          The ratio is measured at $t \sim 3$ for all simulations.
          Generation of fast modes (or $\sim$radial modes) 
           is not very efficient.
          Pluses are for low-$\beta$ cases and diamonds are
          for high-$\beta$ cases.  From Cho \& Lazarian (\cite{CL03}).
        (i) Whistler mode turbulence.
           Initially Fourier modes in the dotted box are excited.
           The contours show energy distribution after
      the turbulence evolves for about one eddy turnover time.
     {}From an upcoming paper.
}
\label{statistics}
\end{center}
\end{figure}

The large scale Alfv\'enic turbulence can interact with protons and
only with high
energy electrons. For a more efficient acceleration or heating of the
electrons one must rely on the cascade of the generated turbulence down
to small
scales.  The
details of this cascade for low beta plasma ($\beta_p \ll 1$) are
discussed in \cite{CL}, where it is demonstrated that the
incompressible Alfv\'en waves
and compressible (slow and fast) waves have distinct scaling relations.
In \cite{GS95} it was stated that incompressible Alfv\'enic modes 
should transfer energy to small scales over a hydrodymic eddy turnover
time $l_{\bot}/v_l$, where $l_{\bot}$ eddy size measured perpendicular
to magnetic field direction. It was proved in \cite{CLV02b} that the
motions perpendicular to magnetic field are identical to hydrodynamic
motions. In other words, while magnetic fields resist bending, they
allow mixing hydrodynamic-type motions. Therefore, the power spectrum of
Alfv\'enic motions perpendicular to magnetic field lines is Kolmogorov-type
(i.e. $v_l\sim l^{1/3}$). Those mixing motions induce waves propagating
along magnetic field lines. The corresponding 
motions involve bending of
magnetic field over scales $l_{\|}$. 
The relation between the bending and mixing motions for an eddy
is given by the condition that the period of the wave propagating
along magnetic field lines is equal to the period of the mixing motions
on a scale that excites this wave $l_{\|}/V_A\sim l_{\bot}/v_l$ (\cite{GS95}).
As the result of this coupling the eddies get  more anisotropic
(i.e. $l_{\bot}\sim l_{\|}^{2/3}$) at small scales, i.e. $l_{\bot}\ll l_{\|}$.
 
Rapid transfer of energy within the Alfv\'enic cascade makes the much slower
non-linear interaction of Alfv\'en and fast modes unimportant. Therefore
the Alfv\'en modes follow the Goldreich-Shridhar scalings even in compressible
media.  
Slow waves in
magnetically dominated plasma plasma move with sound velocity 
$v_s\ll V_A$ and are passively mixed up by much
faster Alfv\'en waves.  Therefore,
similar to the passive scalar, slow modes exhibit the
Goldreich-Shridhar\cite{GS95} scaling/anisotropies. Fast modes in magnetically
dominated plasmas move with the velocity (equal to $V_A$) that does
not depend on the local direction of the magnetic field. Therefore
shearing motions that arise from Alfv\'enic modes modify the fast modes
only marginally. As the result, the fast modes form a distinct
 cascade of their own \cite{CL}. Motions associated with this cascade
are isotropic.

The
transfer of energy to electrons happens through collisionless damping for
compressible mode and through the transformation of the Alfv\'en modes into
the whistler modes. The later process requires more investigation.
The unclear issues are related to (a) the efficiency of the energy transfer
from Alfv\'en modes to whistlers, (b) possible modification of
the Alfv\'en cascade if such a transfer presents a bottleneck effect,
(c) the tranfer of energy from whistlers to protons. While the issues (a)
and (b) are still at the early stages of numerical
investigation, preliminary results
have been obtained for (c). It is clear that
how efficiently whistler turbulence heats protons
depends on the degree of the anisotropy assotiated with whistler modes.
The original whistlers are likely to have the anisotropy similar to that
of the Alfv\'en modes that give rise to them. Results of our 
calculations shown in Fig.~1(i) indicate that the whistler modes injected
with a high degree of anisotropy preserve their anisotropy for sufficiently
long time. This may mean that the energy transfer from whistlers to protons
may not be efficient. Longer integration will provide a more definitive
answer.

\section{Interactions of Turbulence and Particles}

The
resonant interaction of 
energetic particles with MHD turbulence 
has been 
suggested as a mechanism for scattering 
of cosmic rays
and for acceleration of the radiating electrons and protons in many astrophysical plasmas 
(e.g. in solar flares; see \cite{VP99}).
Specifically, the resonance condition is
\( \omega -k_{\parallel }v\mu =n\Omega  \), (\( n=0,\pm 1,2... \)),
where \( \omega  \) is the wave frequency, \( k_{\parallel } \)
is the parallel component of wave vector along the magnetic field,
\( v \) is the particle velocity, \( \mu  \) is the pitch angle
cosine to the magnetic field, \( \Omega  \) is the Larmor frequency
of the particle. Basically there are two main types of resonant interaction:
transit 
and gyroresonance 
.

For the {\it transit acceleration}, the energy exchange corresponds to 
resonance
at \( n=0 \). It is the resonant interaction with parallel magnetic
field perturbations, and therefore only concerning compressible waves. In 
the wave frame, the perturbations are stationary, particles will be 
affected by the magnetic mirror force
\( -(mv_{\perp }^{2}/2B)\nabla _{\parallel }B \). For small amplitude
 waves, particles must be in phase with the wave in order to be reflected 
by the compression. This requires the particles to have component of
velocity $v\mu$ equal to the Alfv\'en velocity $V_A$.
Particles gain energy
in head-on collisions and lose energy in on-tail collisions. Since
the frequency of head-on collisions is greater than that of trailing
collisions, particles will gain a net amount of energy. (\cite{Sch98}).

{\it Gyroresonance} is a resonant interaction between a particle and the
transverse electric field of a wave. It occurs when the Doppler shifted
frequency of the wave in the particle's guiding center rest frame
\( \omega _{gc}=\omega -k_{\parallel }v\mu  \) is a multiple of the
particle gyrofrequency, and the rotating direction of wave electric
vector is the same with the direction for Larmor gyration of the particle.
Thus from this resonance condition, we
know that the most important interaction occurs at 
\( k_{\parallel }=k_{res}\sim \Omega /v_{\parallel }\sim r_{L} \),
the Larmor radius of the high-energy particles.

The quasi-linear theory (QLT) can be used to describe  the interaction 
between particles and MHD waves
(\cite{Sch98}). As we know, GRBs require very strong
background magnetic field \( B_{0} \). Therefore it's very likely
that the perturbation on the resonant scale \( \delta B\ll B_{0} \).
The  diffusion coefficients of the
Fokker-Planck equation that describe
the evolvement of the distribution function of the particles
are determined by the statistical 
properties of the
MHD turbulence in the medium. Frequently, for practical calculations
the MHD turbulence spectrum
is assumed to be isotropic and Kolmogorov. However, our
considerations above testify that this is an erroneous assumption.  
The correlation tensors of the perturbations
for Alfv\'en modes and compressible modes that account for the actual
properties of MHD turbulence have been obtained
in \cite{CLV02b, CL, YL02}.

The calculations in \cite{YL02} that made use of the tensors provided
the scattering efficiency of
anisotropic Alfv\'{e}nic turbulence (see Fig.1a).  We see from 
Figure 2a
that the scattering is substantially suppressed, compared to the
Kolmogorov turbulence that is usually used for scattering calculations (
see also \cite{Chana}).
This happens, first of all,
because most turbulent energy in GS95 turbulence goes to
\( k_{\perp } \) so that there is much
less energy left in the resonance point 
\( k_{\parallel }=(\mu r_{L})^{-1} \).
Furthermore, \( k_{\perp }\gg k_{\parallel } \)
means \( k_{\perp }\gg r_{L}^{-1} \)
so that energetic particles sample many eddies during one gyration
(\cite{Chana}).
This random walk decreases the scattering efficiency by a factor of
\( (\Omega /k_{\perp }v_{\perp })^{1\over 2}=(r_{L}/l_{\perp })^{1\over 2} \),
where \( l_{\perp } \) is the turbulence scale perpendicular to magnetic
field.

Thus the gyroresonance with Alfv\'{e}nic turbulence
is not an effective scattering mechanism 
if turbulence is injected on the large scales, since the
degree of anisotropy increases on smaller scales.
However, if energy is injected isotropically at small scales,
the resulting turbulence would be
more isotropic and scattering will be more efficient. 
Scattering by fast modes can be more efficient since they are 
isotropic.
Yan \& Lazarian \cite{YL02} (see also \cite{LCY} for a review)
performed calculations taking into account
the collisionless damping of fast modes and showed that the scattering 
by fast modes is the dominant scattering mechanism (see Fig.1b). 

\begin{figure}
\begin{center}
\epsfig{file=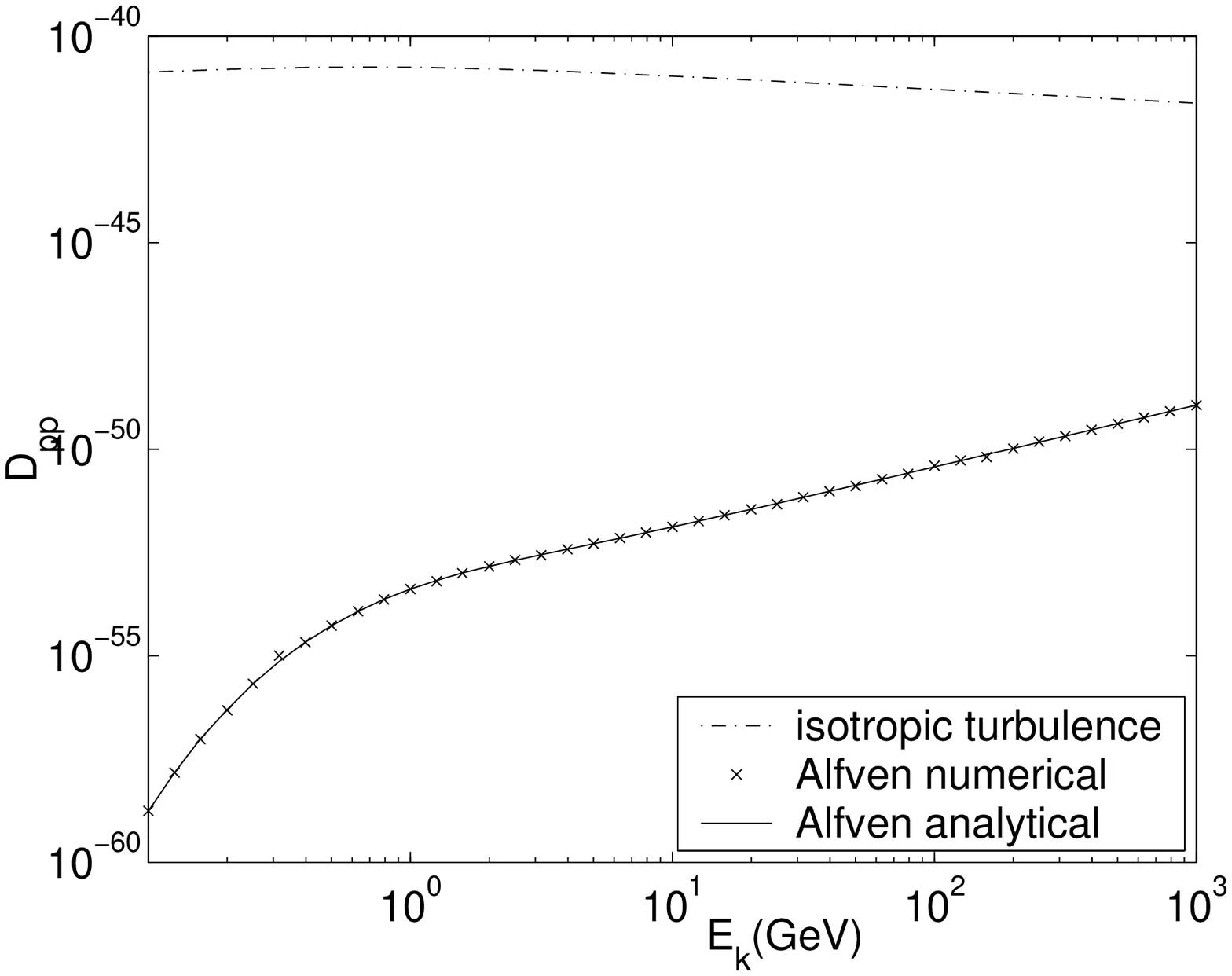,height=1.5in} \hfil
\epsfig{file=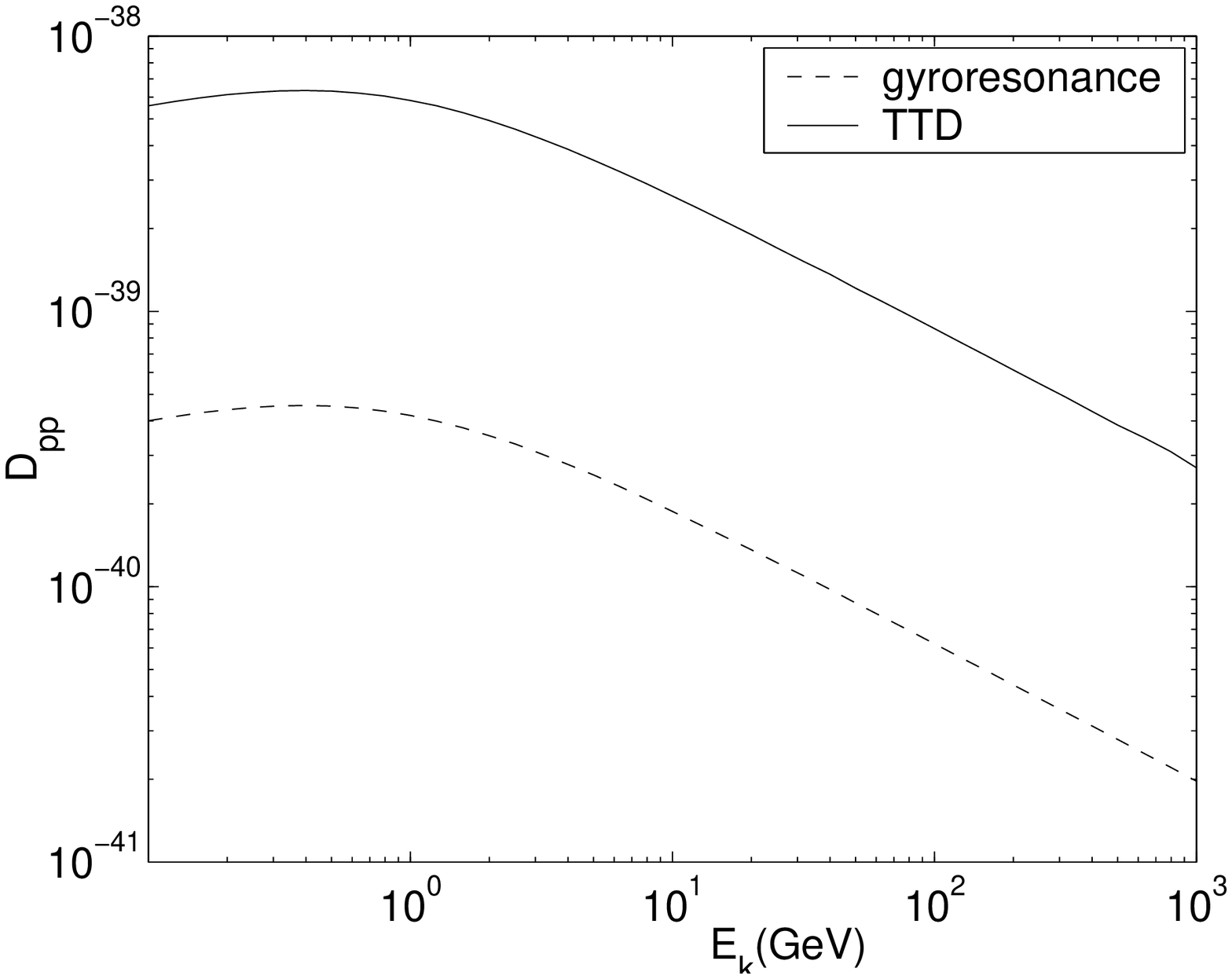,height=1.5in}
\caption{Momentum diffusion \(D_{pp}\) from scattering (a) by 
Alfv\'{e}nic turbulence,
(b) by fast modes. In (a), the dash-dot line refers to the energy 
diffusion in isotropic turbulence. The cross(x) line
represents our result for anisotropic turbulence. In (b), the dashed 
line represents the gyroresonance with fast modes, the solid line are 
the results for transit time acceleration. }
\label{fig:magnet}
\end{center}
\end{figure}

In general, the Fokker-Planck equation is complicated. However, we can get 
an approximate solution for acceleration in certain conditions. 
For instance, if the particle distribution is isotropic, then the acceleration 
process can be described by the so-called {\it diffusion-convection} equation (\cite{Kirk}).
This is applicable at high energies and specifically to {\it cosmic rays}. 
As pointed out above, for GRBs the acceleration is faster than scattering as we pointed 
out earlier, in which case the Fokker-Planck equation can be 
simplified as (\cite{PryP})

\begin{equation}
{\partial f^\mu\over \partial t}+v\mu{\partial f^\mu\over \partial z}={1\over p^2}{\partial \over \partial p}p^2D_{pp}^\mu{\partial f^\mu\over \partial p}+S^\mu,
\end{equation}
where $S^\mu$ is the source term. This expression is similar to the isotropic case except that now everything is a 
function of $\mu$. 
%

Nevertheless, there are still several problems left if we want to apply 
this result to GRBs. The important difference comes from the unusually 
strong magnetic field for GRBs. As mentioned above, the Alfv\'enic velocity
approaches the velocity of light. 
With such a high phase 
speed, the dynamics of the MHD 
turbulence should be taken into account. Instead of the sharp \(\delta\) 
function, we should use the Breit-Wigner-type function 
(\cite{Sch98}, \cite{YL02}) to describe resonance interactions. 
Weeding out the spurious contributions that arise from particle interaction
with large scale magnetic fluctuations is a necessary requirement
for accounting for MHD turbulence dynamics within the QLT \cite{YL02}. 
Additional work is required for describing damping in relativistic plasmas.

\subsection{Towards Self-consistent Model}

A usual assumption for the studies of 
propagation of energetic particles is that the
turbulence properties are given
by external sources. However, it 
is clear that if a substantial part
of the plasma particles is going to be accelerated, this assumption cannot
be true. Instead, the back-reaction of the accelerated particles on the
turbulence is essential. In practical terms that means that a system of
two Fokker-Planck equations, one describing the evolution of the
distribution function of
particles and another describing the evolution of the turbulence energy
should be solved. The coupling of the equations is provided by the fact that
the coefficients of the Fokker-Planck equation describing particle 
acceleration
depend on the turbulence energy at a particular wavenumber, while the
rates of damping in the equation describing turbulence do depend on the
particle distribution (see \cite{Larosa}).

The complication arises from the fact that the 
actual MHD turbulence is anisotropic and this calls for revision of
the earlier studies making use of the recent theoretical
and numerical insights. While the generalization of the particle diffusion
equation is staightforward, the equation that describes the turbulent
cascade requires more care. 
Generalizing expression in \cite{zhou}
we can describe the three-dimensional spectral density 
cascade process as
${\partial W({\bf k}) \over \partial t} =-{\bf \nabla_k}\cdot {\bf F(k)}$, 
where the
flux ${F_i({\bf k})}=-D_{ij} {\bf \nabla}_j W({\bf k})$ and $D_{ij}$ is
the diffusion tensor. 
Here, $\nabla_i\equiv {\partial \over \partial k_i}$ and
summation over repeated indices is assumed. With addition of sources and 
losses the wave transport equation becomes 

\begin{equation}
\label{W}
 {\partial W({\bf k}) \over \partial t}
 = {\dot Q}_p({\bf k}) 
 - \left[ \gamma_e({\bf k}) + \gamma_p({\bf k}) \right] W({\bf k}) 
 +  \nabla_i \left[ D_{ij}
                 \nabla_j W({\bf k}) \right]
 - {W({\bf k}) \over T_{\rm esc}^W}.
\end{equation} 
Here ${\dot Q}_p({\bf k})$ is the rate of wave generation, 
$\gamma_e$ and $\gamma_p$ are the rates of wave damping by the
electrons and protons, ans as for particle transport $T_{\rm esc}^W$ describes 
wave 
leakage from the source region, if any.  
The damping coefficients depend not only on the background
plasma but also on the distribution of the accelerated particles.  This is how
this equation is  coupled to the particle 
transport 
equation (e.g. eq.[1])
, where the diffusion 
coefficients depend on $W({\bf k})$.

As shown in \cite{CL}, the fast mode cascade is isotropic (i.e. 
$D_{ij}$
is a scalar function).
Therefore, we have a one-dimensional spectral energy density distribution;
${\hat W}(k)=4\pi k^2 W(|{\bf k}|)$ and  $D/k^2 \sim c\beta_A k (8\pi {\hat 
W}({\bf k})/B^2)$ gives the rate of the cascade.  
The resultant isotropic case of equation
(\ref{W}) describes an acoustic type cascade of fast waves (see \cite{Larosa}). 

For  Alfv\'enic cascade the losses are negligible till 
the scale of the motions 
($\sim k^{-1}$) reaches much lower values than $L$, possibly till it approaches 
the 
proton Larmor radius.  In this case the wave equation reduces to 
$\nabla_i \left[ D
\nabla_i W({\bf k}) \right] = {\dot Q}_p({\bf k})$, which is independent of the
electron distribution.
In other words, one can
assume that almost all of the injected energy ${\dot Q}_p({\bf k})$ 
reaches the Larmor radius of a proton.  
However, this cascade will yield a non isotropic wave spectrum, 
favoring perpendicular propagation, even if it is
isotropic at injection.  This anisotropy increases with increasing $k$ (or
decreasing scale) approximately as $(kl_0)^{1/3}$, where 
$l_0$ is the energy injection scale (see \cite{CLV02a}).  
At scales near the proton Larmor radius, whistler waves will be
generated.  Quataert \& Gruzinov (\cite{Quat}) speculated that due to the 
anisotropy of the Alfv\'enic turbulence cascade they expect the whistler 
turbulence to be highly
anisotropic.  Our preliminary analysis indicates that whistler turbulence will
eventually become isotropic. 
(see also \cite{goldr}).
Therefore, it can be described by an equation similar to (\ref{W}), where
according to \cite{zhou} the cascade rate is given by $D/k^2\sim 
c\beta_A k(8\pi {\hat W}({\bf
k})/B^2)^{1/2}$.  This means that most of the 
energy from the Alfv\'enic cascade goes into the
whistler mode, which can accelerate electrons of essentially all energies and
pitch angles very efficiently (\cite{HPetro, PryP}).

\section{Gamma Ray Bursts and Magnetic Reconnection}

An 
alternative model
for energizing GRBs, 
that we heard about at the conference, is reconnection,
i.e. the process of magnetic field annihilation.
This willrequire a rapid rate of energy release.
Without elaborating on the details, 
we briefly discuss 
the conditions under which
magnetic reconnection can provide rapid bursts of energy release.

In the existing models of Gamma Ray bursts magnetic field play important
role. If magnetic field is sufficiently high its energy may be
sufficient to feed the Gamma Ray burst itself (see other papers in the
volume that explore this possibility). For this purpose a mechanism
of a fast release of energy stored in magnetic field is necessary. The
problem that one encounters here is similar to the one in
solar flare research where a relatively slow accumulation of the oppositely
directed flux is required to be followed by  a catastrophic
release of energy on a short time scale. Therefore to understand
the reconnection one should understand why the reconnection can
be both fast and slow.

It is trivial to understand why reconnection can be slow in most astrophysical
environments. Indeed, the ratio of the advection term to the diffusion term
in the induction equation is given by the so-called Lundquist number ${\cal R}_L=(V_A L/\eta)$,
where $V_A$ is the Alfv\'en velocity, $L$ ($>10^{13}$~cm) 
is the scale of the system and $\eta=c^2/(4\pi \sigma)$ is
magnetic diffusivity. For plasma of temperature $T$, $\eta=(10^{9}/T)^{-3/2}$~cm$^2$ s$^{-1}$
and so
typically ${\cal R}_L\gg 1$ and therefore the magnetic field diffusion
term is negligible compared to  advection.

The literature on magnetic reconnection is rich and vast (see, for
example, \cite{Bisk93} and references therein). We start by discussing
a robust scheme proposed by Sweet and Parker  (\cite{Parker,Sweet}).
In this scheme oppositely directed magnetic fields are brought
into contact over a region of size $L_x$ (see Fig.~2). The diffusion of
magnetic field takes place over the vertical scale $\Delta$ which is
related to the Ohmic diffusivity by $\eta\approx V_r \Delta$,
where $V_r$ is the velocity at which magnetic field lines can get into
contact with each other and reconnect. Given some fixed $\eta$ one may naively
hope to obtain fast reconnection by decreasing $\Delta$. However, this
is not possible. An additional constraint posed by mass conservation
must be satisfied. The plasma initially entrained on the magnetic field
lines must be removed from the reconnection zone. In the Sweet-Parker
scheme this means a bulk outflow through a layer with a thickness of
$\Delta$.  In other words the entrained mass
must be ejected, i.e. $\rho V_r L_x = \rho' V_A \Delta$,
where it is assumed that the outflow occurs at the Alfv\'en velocity.
If we ignore the effects of compressibility $\rho=\rho'$ and the
resulting reconnection velocity allowed by Ohmic diffusivity
and the mass constraint is $V_r\approx V_A {\cal R}_L^{-1/2}$,
i.e. very slow. Surely such reconnection cannot explain neither
solar flares nor GRBs.

\subsection{Fast Reconnection}

So far, attempts to explain fast reconnection have not been
supported by subsequent studies.  The `X-point' model of Petschek
(\cite{Pets}) collapses to a Sweet-Parker geometry after a short time
(\cite{Bisk, Wang}) and
recent plasma reconnection experiments (see \cite{Hsu,
Yamada}) show flat Sweet-Parker type current sheets.
While collisionless effects can broaden the current sheet
to roughly the ion skin depth (\cite{Shay1, Shay2, Shay99}), this, 
by itself, may not produce
fast reconnection speeds in astrophysical contexts. For instance,
such studies have not demonstrated  the possibility of
fast reconnection in the presence of the large scale forces
acting to produce a large scale current sheet. Priest and Forbes 
(\cite{Priest})
have stressed that even if fast Petschek
reconnection is possible it will still be necessary to
demonstrate ``that it will apply to the turbulent MHD regime''.
If magnetic fields
are turbulent the boundary conditions for the current sheets
are changing
stochastically.
On the other hand,  boundary conditions need to be fine tuned for a
Petschek reconnection scheme (see \cite{Priest}).
\begin{figure}
\begin{center}
\epsfig{file=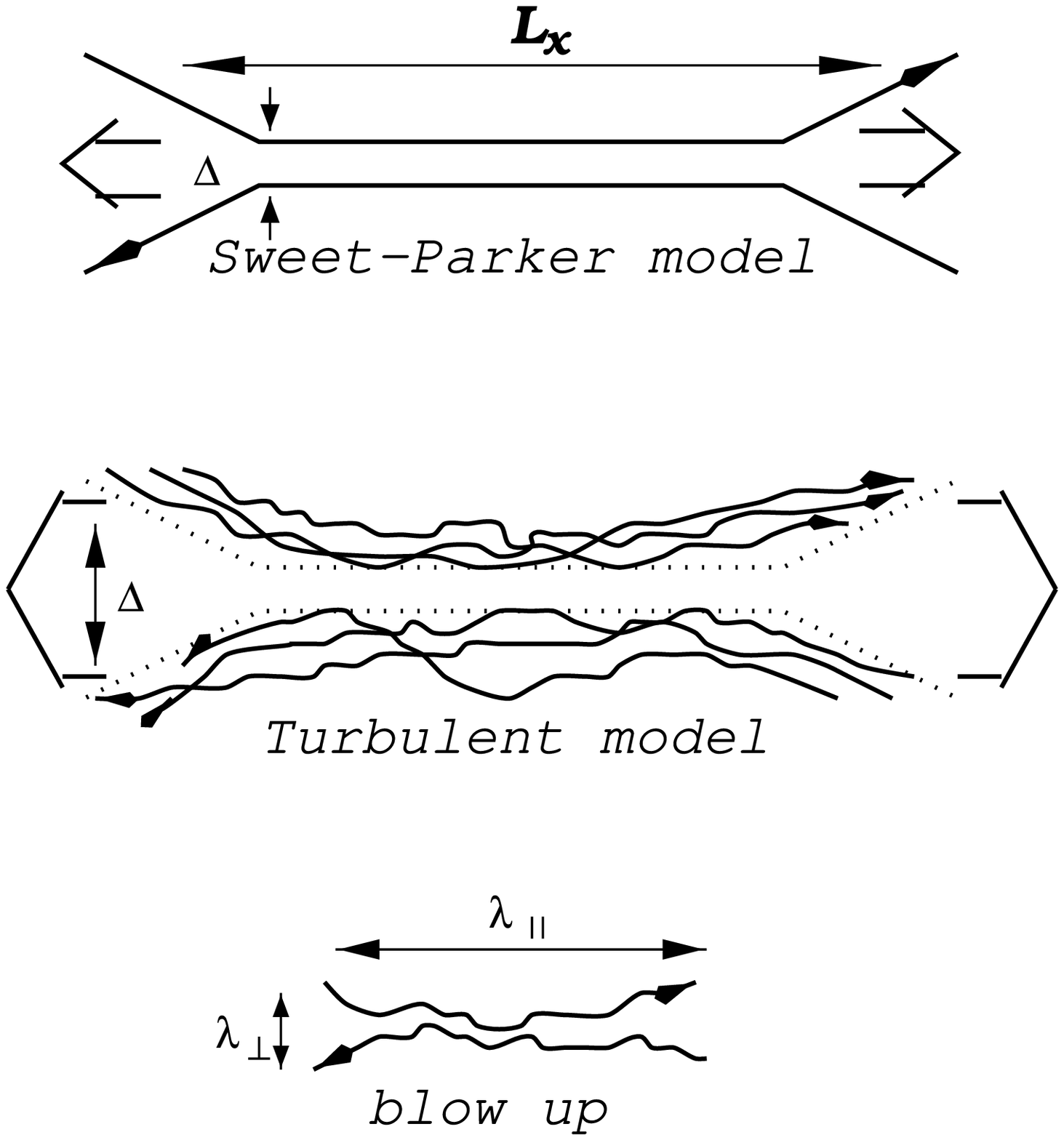,height=2in} \hfil
\reflectbox{\rotatebox{90}{\reflectbox{\epsfig{file=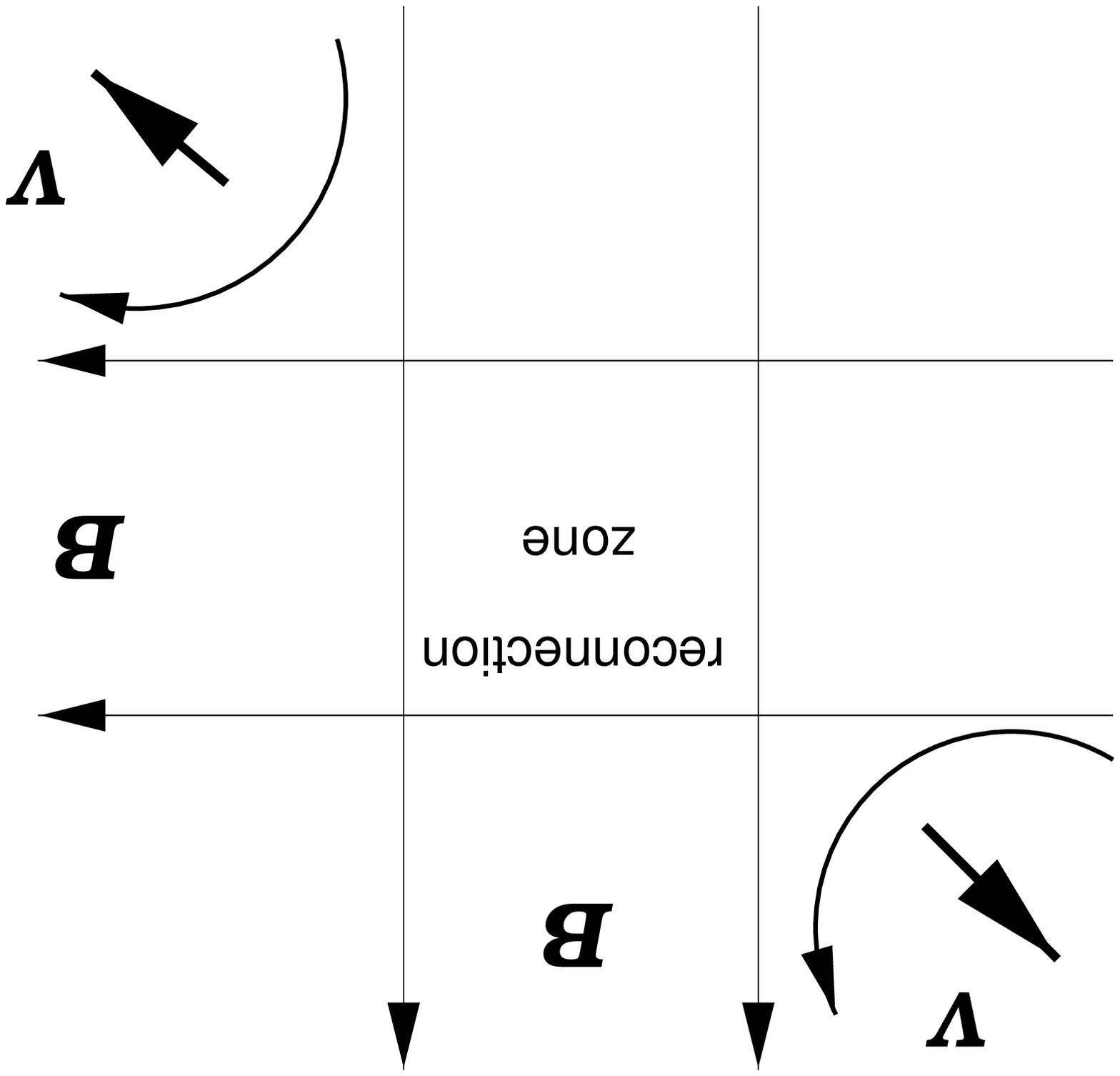,height=2.in}}}}
\caption{{\it Right Panel}. 
Upper plot: Sweet-Parker scheme of reconnection. Middle plot:
the new scheme of reconnection that accounts for field stochasticity,
lower plot: a blow up of the contact region.
Thick arrows depict outflows of plasma. {\it Left Panel}. The stochastic
reconnection scheme requires 3D topology. In 2D the magnetic field
lines cannot independently enter the reconnection region and this 
limits the reconnection speed. In the figure the magnetic fluxes
intersect each other at 90 degrees.
}
\label{fig:magnet1}
\end{center}
\end{figure}

A number of researchers have claimed that turbulence may accelerate
reconnection. For instance, a turbulence-related concept of
hyper-resistivity was put forward by Strauss (\cite{Strauss}). He correctly
pointed out that the turbulence driven by current sheet instabilities may
broaden
the current sheets compared to the Sweet-Parker estimate. However,
such instabilities, i.e. the tearing mode instability,
do not allow us to evade the constraints on the global
plasma flow that lead to slow reconnection speeds, a point which
has been demonstrated numerically (\cite{Matt})
and analytically (Lazarian \& Vishniac \cite{LV99} 1999, hereafter LV99).

Nevertheless, a further analysis of the stochastic reconnection resulted in
a new model of fast reconnection that we shall briefly describe below.
A detailed discussion is given in LV99.

MHD turbulence guarantees the presence of a stochastic field component
properties of which depend on the admixture of compressible
and incompressible modes (\cite{CL, CLV02b}).
We consider the case in which there exists a large scale,
well-ordered magnetic field, of the kind that is normally used as
a starting point for discussions of reconnection.  This field may,
or may not, be ordered on the largest conceivable scales.  However,
we will consider scales smaller than the typical radius of curvature
of the magnetic field lines, or alternatively, scales below the peak
in the power spectrum of the magnetic field, so that the direction
of the unperturbed magnetic field is a reasonably well defined concept.
In addition, we expect that the field has some small scale `wandering' of
the field lines.  On any given scale the typical angle by which field
lines differ from their neighbors is $\delta \phi\ll1$, and this angle persists
for a distance along the field lines $\lambda_{\|}$ with
a correlation distance $\lambda_{\perp}$ across the field lines (see Fig.~2).

The modification of the mass conservation constraint in the presence of
a stochastic magnetic field component
is self-evident. Instead of being squeezed from a layer whose
width is determined by Ohmic diffusion, the plasma may diffuse
through a much broader layer, $L_y\sim \langle y^2\rangle^{1/2}$ (see Fig.~2),
determined by the diffusion of magnetic field lines.  This suggests
an upper limit on the reconnection speed of
$\sim V_A (\langle y^2\rangle^{1/2}/L_x)$.
This will be the actual speed of reconnection
the progress of reconnection in the current sheet itself does not
impose a smaller limit. The value of
$\langle y^2\rangle^{1/2}$ can be determined once a particular model
of turbulence is adopted, but it is obvious from the very beginning
that this value is determined by field wandering rather than Ohmic
diffusion as in the Sweet-Parker case.

What about limits on the speed of reconnection that arise from
considering the structure of the current sheet?
In the presence of a stochastic field component, magnetic reconnection
dissipates field lines not over their  entire length $\sim L_x$ but only over
a scale $\lambda_{\|}\ll L_x$ (see Fig.~2), which
is the scale over which magnetic field line deviates from its original
direction by the thickness of the Ohmic diffusion layer $\lambda_{\perp}^{-1}
\approx \eta/V_{rec, local}$. If the angle $\phi$ of field deviation
does not depend on the scale, the local
reconnection velocity would be $\sim V_A \phi$ and would not depend
on resistivity. In LV99 we claimed that $\phi$ does depend on scale.
Therefore the {\it local}
reconnection rate $V_{rec, local}$ is given by the usual Sweet-Parker formula
but with $\lambda_{\|}$ instead of $L_x$, i.e. $V_{rec, local}\approx V_A
(V_A\lambda_{\|}/\eta)^{-1/2}$.
It is obvious from Fig.~2 that $\sim L_x/\lambda_{\|}$ magnetic field
lines will undergo reconnection simultaneously (compared to a one by one
line reconnection process for
the Sweet-Parker scheme). Therefore the overall reconnection rate
may be as large as
$V_{rec, global}\approx V_A (L_x/\lambda_{\|})(V_A\lambda_{\|}/\eta)^{-1/2}$.
Whether or not this limit is important depends on
the value of $\lambda_{\|}$.
The relevant values of $\lambda_{\|}$ and $\langle y^2\rangle^{1/2}$
depend on the magnetic field statistics. This
calculation was performed in LV99 using the 
Goldreich-Sridhar\cite{GS95} model
of MHD turbulence.
The upper limit on $V_{rec,global}$ was greater
than $V_A$, so that the diffusive wandering of field lines imposed
the relevant limit on reconnection speeds.
Thus
\begin{equation}
V_{r, up}=V_A \min\left[\left({L_x\over l}\right)^{\frac{1}{2}}
\left({l\over L_x}\right)^{\frac{1}{2}}\right]
\left({v_l\over V_A}\right)^{2},
\label{main}
\end{equation}
where $l$ and $v_l$ are the energy injection scale and
turbulent velocity at this scale respectively.
In LV99 we also considered other processes that can impede
reconnection and find that they are less restrictive. For
instance, the tangle of reconnection field lines crossing the
current sheet will need to reconnect repeatedly before individual
flux elements can leave the current sheet behind.  The rate at which
this occurs can be estimated by assuming that it constitutes the
real bottleneck in reconnection events, and then analyzing each
flux element reconnection as part of a self-similar system of
such events.  This turns out to limit reconnection to speeds less
than $V_A$, which is obviously true regardless.  As the result we
showed in LV99 that (\ref{main}) is not only an
upper limit, but is the best estimate of the speed of reconnection.

\subsection{Flares of Reconnection}

Evidently the reconnection rate given by eq.~(\ref{main}) is
large, i.e. of the order of maximal possible velocity, which is
$V_A$. Can the reconnection be slow at all? Naturally,
when turbulence is negligible, i.e. $v_l\rightarrow 0$, the
field line wandering is limited to the Sweet-Parker current sheet
and the Sweet-Parker reconnection scheme takes over.

We also note that observations of solar flares
seem to show that reconnection events start from some limited
volume and spread as though
a chain reaction from the initial reconnection region initiated
a dramatic change in the magnetic field properties. Indeed, a
solar flare happens as if the
resistivity of plasma were increasing dramatically as plasma
turbulence grows (see \cite{Dere} and references therein).
In the LV99 picture this is a consequence of
the increased stochasticity of the field lines rather than
any change in the local resistivity.  The change in magnetic
field topology that follows localized reconnection provides
the energy necessary to feed a turbulent cascade in neighboring
regions.  This kind of nonlinear feedback is also seen in the
geomagnetic tail, where it has prompted the suggestion that
reconnection is mediated by some kind of nonlinear instability
built around modes with very small $k_{\|}$ (cf. \cite{Chang} and
references therein).  The most detailed exploration
of nonlinear feedback can be found in the work of Matthaeus and
Lamkin (\cite{Matt86}), who showed that instabilities in narrow current
sheets can sustain broadband turbulence in two dimensional
simulations.
Although the LV99 model is quite different, and relies on the
three dimensional wandering of field lines to sustain fast
reconnection,  we note that the concept
of a self-excited disturbance does carry over and may describe
the evolution of reconnection between volumes with initially
smooth magnetic fields.

To the best of our knowledge, there are no detaled calculations
for GRBs
using the stochastic reconnection scheme. 
Further research is necessary to show whether this is a viable
alternative to the more traditional models of GRBs. Most
of the energy released via stochastic reconnection goes into MHD
turbulence. Therefore the particle acceleration will happen according to
the model described in sections 4-5.

\section{Summary}

\begin{itemize}
\item Whether the GRB arises from the interaction of relativistic shocks
or due to magnetic reconnection MHD turbulence is likely to
play an essential role in transfering the energy to electrons.
\item Recent understanding of fundamentals of MHD turbulence as well
as particle acceleration by compressible and Alf\'enic modes
of the turbulence allows quantitative description of GRBs.
\item If magnetic 
reconnection plays a role in GRBs, the stochastic reconnection
is the primary candiate to produce bursts.
\end{itemize}

{\bf Acknowledgments}. The work of AL, HY and JC is supported by NSF
grant AST0125544.

\def\Discussion{
\setlength{\parskip}{0.3cm}\setlength{\parindent}{0.0cm}
     \bigskip\bigskip      {\Large {\bf Discussion}} \bigskip}
\def\speaker#1{{\bf #1:}\ }
\def\endDiscussion{}

\endDiscussion

\end{document}